\documentclass{my-ws-p10x7}
\begin{document}

\title{
\hspace*{-0.5cm}
{\rm\mbox{(hep-ph/0007nnn, 31.~July 2000)}}\\[0.7cm]
\boldmath Implications of a Low $\sin2\beta$:\unboldmath\\
A Strategy for Exploring New Flavor Physics}

\author{Alexander L. Kagan}
\address{Department of Physics, University of Cincinnati\\
Cincinnati, OH 45221, USA}

\author{Matthias Neubert}
\address{Newman Laboratory of Nuclear Studies, Cornell University\\ 
Ithaca, NY 14853, USA}  

\twocolumn[\maketitle\abstract{
We explore the would-be consequences of a low value of the 
CP-violating phase $\sin2\beta_{\psi K}$. The importance of a reference 
triangle obtained from measurements that are independent of 
$B$--$\bar B$ and $K$--$\bar K$ mixing is stressed. It can be used to 
extract separately potential New Physics contributions to mixing in 
the $B_d$, $B_s$ and $K$ systems. We discuss several constructions of 
this triangle, which will be feasible in the near future. The discrete 
ambiguity is at most two-fold and eventually can be completely 
removed. Simultaneously, it will be possible to probe for New Physics 
in loop-dominated rare decays.}]

\section{Introduction}

One of the highlights of the ongoing 
ICHEP2000 Conference in Osaka
has been the presentation of first results on CP violation in 
$B_d$--$\bar B_d$ mixing by the BaBar and Belle Collaborations 
\cite{Osaka}. The reported values obtained from the time-dependent
CP asymmetry in $B\to J/\psi\,K_S$ decays,
\begin{equation}
   \sin2\beta_{\psi K} = \cases{
    0.12\pm 0.37\pm 0.09 \,; & BaBar, \vspace{0.1cm} \cr
        0.45_{\,-0.44-0.09}^{\,+0.43+0.07} \,; & Belle, \cr}
\end{equation}
are smaller than the previous best measurement 
$\sin2\beta_{\psi K}=0.79_{-0.44}^{+0.41}$ by the CDF Collaboration 
\cite{CDF}. 
They are also smaller than the value $\sin2\beta=0.75\pm 0.06$ 
obtained from a recent global analysis of the unitarity 
triangle~\cite{CPRS00}. Although there is at present no statistically 
significant discrepancy, it is interesting to explore the implications 
of a measurement of $\sin2\beta_{\psi K}$ that would be inconsistent 
with the results of the global analysis. Under the assumption that 
there is no 
CP-violating New Physics in $b\to c\bar c s$ transitions (which is 
supported by the strong experimental bound on direct CP violation in 
$B^\pm\to J/\psi\,K^\pm$ decays reported by the CLEO Collaboration 
\cite{CLEOCP}) this would imply the existence of New Physics in 
$B_d$--$\bar B_d$ mixing. (New Physics in $K$--$\bar K$ mixing could 
not account for such a discrepancy, because of the minor impact of 
$|\epsilon_K|$ on the global analysis.) The measured phase 
$2\beta_{\psi K}$ would then be the $B_d$--$\bar B_d$ mixing phase
$2\phi_d$, which would differ from the CKM phase $2\beta$ because of
New Physics. In such an event, it is likely that New Physics would 
also play a role in $B_s$--$\bar B_s$ and $K$-$\bar K$ mixing. 

The purpose of this Letter is to point out a strategy which provides 
a systematic exploration of the new flavor physics in this case.
This strategy is different from the conventional route 
pursued at the $B$ factories, in which the main focus 
is on measurements that are sensitive to $B$--$\bar B$ mixing, mainly 
because CP violation in the interference of mixing and decay can
sometimes be interpreted without encountering large hadronic 
uncertainties. If there is New Physics in mixing, then the standard
triangle obtained by combining information on $|V_{ub}|$ from 
semileptonic $B$ decay, $\Delta m_{d,s}$ from $B_{d,s}$--$\bar B_{d,s}$ 
mixing, and $|\epsilon_K|$ from $K$--$\bar K$ mixing does not
agree with the true CKM triangle, and forcing it to close (as is done
in the standard analysis) gives wrong results for the angles 
$\gamma=\mbox{arg}(V_{ub}^*)$ and $\beta=-\mbox{arg}(V_{td})$. 
(We use the standard phase conventions; otherwise 
$\gamma=\mbox{arg}[-(V_{ub}^* V_{ud})/(V_{cb}^* V_{cd})]$ and 
$\beta=\mbox{arg}[(V_{tb}^* V_{td})/(V_{cb}^* V_{cd})]$.) 

In the absence of a reliable way to measure the magnitude and phase 
of $V_{td}$ in $B$ decays, it is important to base studies 
of the CKM matrix on a {\em reference triangle\/} obtained
exclusively from measurements independent of particle--antiparticle
mixing \cite{GKOT95,CKLN97,GNW97,BEN99}. 
In the $B$ system, this triangle is constructed from the 
measurement of the magnitude and phase of $V_{ub}$. (The use of 
$\gamma=\mbox{arg}(V_{ub}^*)$ replaces the use of $|V_{td}|$, 
determined from $B$--$\bar B$ mixing, in the standard analysis.) 
Separate comparisons of particle--antiparticle mixing measurements 
in the $B_d$, $B_s$ and $K$ systems with information obtained from 
the reference triangle will allow extraction of the magnitude and
phase of New Physics contributions to the mixing amplitudes. At a 
later stage, comparison of different reference triangle 
constructions can provide information about potential New Physics 
effects in the decay amplitudes, not related to mixing. 

We stress that the reference triangle approach should be pursued
regardless of whether or not the $\sin2\beta_{\psi K}$ measurements
are consistent with the global analysis of the unitarity triangle.
Agreement within errors could be accidental and would not exclude 
the possibility of large New Physics contributions in 
$B_d$--$\bar B_d$ mixing.
Proposals similar in spirit to ours have been discussed in
the past. However, their feasibility is limited by their 
reliance on methods for extracting $\gamma$ that are extremely 
difficult, and are plagued by multiple 
discrete ambiguities \cite{GKOT95,GNW97}. In the past two years,
however, several strategies have been proposed that will allow a
determination of $\gamma$ in the near future, without discrete 
ambiguities and with controlled theoretical uncertainties.

Our analysis is based on the following standard assumptions, which
hold true for a vast class of extensions of the Standard Model (for
a discussion, see e.g.\ Ref.~\cite{BEN99}):

i) The determination of the CKM elements $|V_{us}|$, $|V_{cb}|$ and
$|V_{ub}|$ from semileptonic decays is not affected by New Physics.

ii) The 3-generation CKM matrix is unitary.

iii) There are no (or negligibly small) New Physics effects in 
decays which in the Standard Model are dominated by tree 
topologies.

An experimental test for non-standard contributions 
in the semileptonic $b\to u\,l\,\nu$ transition could be 
performed by comparing the values of $|V_{ub}|$ extracted from the
exclusive $B\to\pi\,l\,\nu$ and $B\to\rho\,l\,\nu$ decays, and the 
inclusive $B\to X_u\,l\,\nu$ decays. (An analogous test for
$b\to c\,l\,\nu$ decays has been proposed in Ref.~\cite{Volo97}.) 
Tests of the unitarity of the CKM matrix will be discussed in 
Section~\ref{sec:boxes}.

\section{The reference triangle}
\label{sec:RT}

Disregarding all information obtained from mixing measurements, not
much is known about the Wolfenstein parameters $(\bar\rho,\bar\eta)$
determining the unitarity triangle. 
The magnitude of $|V_{ub}|$ measured in semileptonic $B$ decay fixes 
$R_b=|(V_{ub}^* V_{ud})/(V_{cb}^* V_{cd})|
=\sqrt{\bar\rho^2+\bar\eta^2}$, corresponding to a circle 
centered at the origin in the $(\bar\rho,\bar\eta)$ plane. 
At present $|V_{ub}|$ is known with a precision of about 20\%.
A reduction of the uncertainty to the 10\% level appears realistic 
within a few years. The phase $\gamma$ defining the orientation of 
the triangle ($\sin\gamma=\bar\eta/R_b$) is currently unknown. 

\subsection{Reference triangles from $B$ decays}

In the very near future, ratios of CP-averaged $B\to\pi K$ and 
$B\to\pi\pi$ branching ratios can be used to extract $\cos\gamma$ 
using many different strategies. A method based on flavor symmetries,
using little theory input, has been described in \cite{Mont99}. 
It makes use of two experimentally determined rate ratios ($R_*$ and 
$\bar\varepsilon_{3/2}$) and the theoretical prediction\cite{BBNS99} 
that the relevant strong-interaction phase cannot be very large.
Alternatively, it has been argued recently that in the heavy-quark 
limit most two-body hadronic $B$ decays admit a model-independent 
theoretical description based on a QCD factorization formula 
\cite{BBNS99}. Predictions for the $B\to\pi K,\pi\pi$ decay-rate 
ratios as a function of $\cos\gamma$ have been obtained (including 
the leading power corrections in $1/m_b$) \cite{BBNS00}. The 
combination of several independent determinations of $\cos\gamma$ 
from these modes will fix $\gamma$, up to a sign ambiguity 
$\gamma\to-\gamma$, with reasonable precision. (We define all weak 
phases to lie between $-180^\circ$ and $180^\circ$.) We believe that 
an uncertainty of $\Delta\gamma=25^\circ$ will be attainable in the 
near future.

Once the $B^\pm\to(\pi K)^\pm$ and $B^\pm\to\pi^\pm\pi^0$ decay 
rates are known with higher precision, $\gamma$ can be determined 
with minimal theory input (up to discrete ambiguities) using the 
method of Ref.~\cite{NR98}. Here, in addition to CP-averaged decay 
rates, information about some direct CP asymmetries is added. 
Ultimately, this will reduce the 
theoretical uncertainty to a level of $10^\circ$ or less. When 
supplemented with theoretical information on the strong-interaction
phase this method can be used to completely eliminate the discrete 
ambiguities (for a detailed discussion, see Ref.~\cite{Mont99}). 

\begin{figure}
\epsfxsize=190pt
\centerline{\epsffile{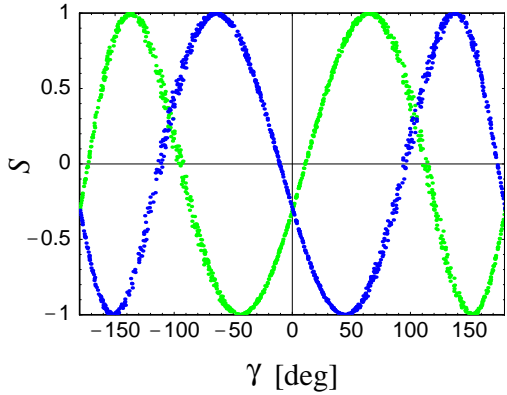}}
\caption{Determination of $\gamma$ from the mixing-induced CP 
asymmetry in $B\to \pi^+\pi^-$ decays, assuming $\sin2\phi_d=0.3$. 
The dark band refers to $2\phi_d\simeq 17.5^\circ$, the light one 
to $2\phi_d=162.5^\circ$.}
\label{fig:gampipi}
\vspace*{-0.5cm}
\end{figure}

The QCD factorization approach can also be used to calculate the
penguin-to-tree ratio in $B\to\pi^+\pi^-$ decays, thereby turning 
a measurement of the mixing-induced CP violation into a determination 
of $\gamma$ without the need for an (impractical) isospin analysis 
\cite{BBNS00}. In Figure~\ref{fig:gampipi} we show
the coefficient $S$ of $-\sin(\Delta m_d\,t)$ in the time-dependent
CP asymmetry as a function of $\gamma$, assuming $\sin 2\phi_d=0.3$ 
for the $B$--$\bar B$ mixing phase measured in $B\to J/\psi\,K_S$. 
A measurement of $\sin2\phi_d$ determines the phase $2\phi_d$ up to
a two-fold discrete ambiguity, $2\phi_d^{(1)}+2\phi_d^{(2)}=\pi$ 
mod $2\pi$. The width of the bands reflects the 
theoretical uncertainty. The eight-fold ambiguity can be reduced to 
a four-fold one, in principle, by measuring the direct CP asymmetry 
in this decay. Alternatively, using a Dalitz-plot analysis of the 
$B\to\rho\pi$ decay amplitudes one could determine $\sin(2\phi_d+2\gamma)$
with small hadronic uncertainties \cite{QS93}.

\begin{figure}
\epsfxsize=190pt
\centerline{\epsffile{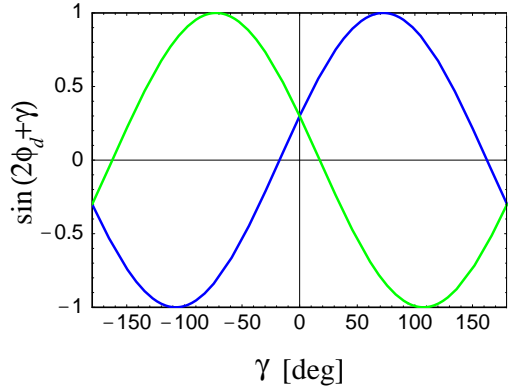}}
\caption{Determination of $\gamma$ from the measurement of 
$\sin(2\phi_d+\gamma)$ in $B\to D^{(*)\pm}\pi^\mp$ decays, assuming 
$\sin2\phi_d=0.3$. The dark curve refers to 
$2\phi_d\simeq 17.5^\circ$, the light one to $2\phi_d=162.5^\circ$.}
\label{fig:gamDpi}
\vspace*{-0.5cm}
\end{figure}

On a much longer time-scale, it will be possible to obtain 
information on $\gamma$ (again up to discrete ambiguities) using 
only decays mediated by tree topologies in the Standard Model. 
Examples are the determination of $\sin(2\phi_d+\gamma)$ from 
$B\to D^{(*)\pm}\pi^\mp$ decays \cite{DS88}, and the extraction of
$\gamma$ from $B\to D K$ decays \cite{ADS97}. This last method, in 
particular, will require very large data samples. Other methods 
make use of $B_s$-meson decays accessible at future $B$ factories 
at hadron colliders \cite{LHCb}. In Figure~\ref{fig:gamDpi} we show 
as an example the information obtainable from a determination of 
$\sin(2\phi_d+\gamma)$. Combining this with the information derived 
from a measurement of the $B\to\pi^+\pi^-$ CP asymmetry (see 
Figure~\ref{fig:gampipi}), a unique pair of solutions 
$(\gamma,2\phi_d^{(1)})$ and $(-\gamma,2\phi_d^{(2)})$ is obtained. 
We stress that combining any of the measurements sensitive to
$B_d$--$\bar B_d$ mixing described above with a determination of 
$\gamma$ (including its sign) from $B\to\pi K$ would remove the 
discrete ambiguity in the $B_d$ mixing phase $2\phi_d$.

Up to now we have assumed that the various determinations of the
reference triangle from $B$ decays are consistent with each other. 
This assumption will have to be tested as the data become 
increasingly precise. In Section~\ref{sec:boxes} we will discuss 
how {\em differences\/} between these constructions would provide 
information about New Physics in $B$ decays rather than 
$B_d$--$\bar B_d$ mixing. 

\subsection{Reference triangle from $K$ decays}

An independent reference triangle can be constructed from 
measurements of very rare kaon decays. The branching ratios for the
decays $K^+\to\pi^+\nu\bar\nu$ and $K_L\to\pi^0\nu\bar\nu$ measure
$|V_{ts}^* V_{td}|$ and $|\mbox{Im}(V_{ts}^* V_{td})|$, 
respectively, and thereby determine 
$R_t=\sqrt{(1-\bar\rho)^2+\bar\eta^2}$ and $|\eta|$ independently 
of $K$--$\bar K$ mixing \cite{BBL96}. This provides a reference 
triangle up to a four-fold discrete ambiguity. Dedicated experiments 
would be necessary to measure $R_t$ and $|\eta|$ with useful 
precision.
In Section~\ref{sec:boxes} we will discuss what can be learned from
the comparison of the kaon reference triangle with the $B$-meson
triangle(s).

\section{Exploring New Physics}

\begin{figure}
\epsfxsize=190pt
\centerline{\epsffile{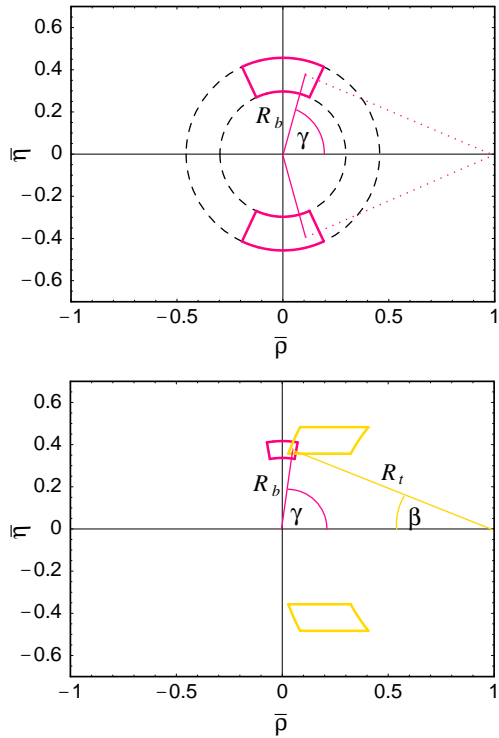}}
\caption{Illustrative examples of reference triangle constructions in 
the near and long-term future. 
(a) $B$-decay triangle with two-fold ambiguity, assuming uncertainties 
of 20\% in $|V_{ub}/V_{cb}|$ and $\pm 25^\circ$ in $\gamma$ 
(near-term). The dashed circles correspond to the measurement of 
$|V_{ub}/V_{cb}|$. 
(b) $B$-decay triangle (dark) with no ambiguity, assuming 
uncertainties of 10\% in $|V_{ub}/V_{cb}|$ and $\pm 10^\circ$ in 
$\gamma$, and $K$-decay triangle (light) with four-fold ambiguity, 
assuming 15\% uncertainties in $R_t$ and $|\eta|$ (long-term). The two
solutions for the kaon triangle obtained with $\bar\rho\to 2-\bar\rho$ 
are not shown.}
\label{fig:RT}
\vspace*{-0.5cm}
\end{figure}

Once the reference triangle is known, we can use it to explore New 
Physics contributions to $B$--$\bar B$ and $K$--$\bar K$ mixing. 
Measurement of $R_b$ and $\gamma$ fix the coordinates $\bar\rho$ and 
$\bar\eta$, which in turn determine the other side $R_t$ and the true 
angle $\beta$ of the reference triangle via
\begin{eqnarray}\label{Rt}
   R_t &=& \sqrt{(1-\bar\rho)^2+\bar\eta^2} \,, \nonumber\\
  \sin\beta &=& \frac{\bar\eta}{R_t} \,, \quad
   \cos\beta = \frac{1-\bar\rho}{R_t} \,.
\end{eqnarray}
In Figure~\ref{fig:RT}, we illustrate what the situation may look
like both in the near and long-term future. We assume that in the
near future $\gamma$ will be known only up to a sign ambiguity (from
measurements of CP-averaged decay rates), which will be resolved 
after several more years of running at the $B$ factories (when certain 
CP asymmetries in rare decays will have been measured). 

\subsection{$B_d$--$\bar B_d$ mixing}

We now discuss how one can systematically study New Physics effects
in the $B_d$--$\bar B_d$ mixing amplitude $M_{12}$ by confronting
measurements of the mass difference $\Delta m_d=2|M_{12}|$ (or 
$x_d=\Delta m_d\,\tau_B$) and of the mixing phase $\sin2\phi_d$ with 
the reference triangle. Our approach is very similar to the one 
discussed in Ref.~\cite{GKOT95}. Using
(\ref{Rt}) we construct the complex quantity $R_t^2\,e^{-2i\beta}$
with the {\em true\/} CKM phase $\beta$. Up to a constant $C_B$, this
quantity determines the Standard Model contribution to the mixing
amplitude: $M_{12}^{\rm SM}=C_B\,R_t^2\,e^{-2i\beta}$, where 
\cite{BBL96}
\begin{eqnarray}
   C_B &=& \frac{G_F^2}{12\pi^2}\,\eta_B\,m_B\,m_W^2\,S_0(x_t)\, 
    B_B f_B^2 \nonumber\\
   &\simeq& 0.24\,\mbox{ps}^{-1} \times
    \frac{B_B f_B^2}{(0.2\,\mbox{GeV})^2} \,.
\end{eqnarray}
The main uncertainty in this result comes from the hadronic matrix
element parameterized by the product $B_B f_B^2$. It is therefore 
convenient to focus on the ratio $M_{12}/C_B$, which in the Standard 
Model is given only in terms of CKM parameters: $M_{12}^{\rm SM}/C_B
=[(1-\bar\rho)^2-\bar\eta^2)]-2i\bar\eta(1-\bar\rho)$. (If New Physics 
does not induce operators with non-standard Dirac structure, the ratio 
$M_{12}/C_B$ remains free of hadronic uncertainties.) The 
experimental value of the mixing amplitude is given by 
$M_{12}^{\rm exp}/C_B=(\Delta m_d/2 C_B)\,e^{-2i\phi_d}$. If the 
mixing phase is determined from the $\sin2\phi_d$ measurement in 
$B\to J/\psi\,K_S$ decays alone, then $e^{-2i\phi_d}$ has a two-fold 
discrete ambiguity. In the previous section we have discussed how this
ambiguity may eventually be resolved by using data on CP violation in
$B$ decays. The difference 
$M_{12}^{\rm NP}=M_{12}^{\rm exp}-M_{12}^{\rm SM}$ is the New Physics 
contribution to the mixing amplitude.

\begin{figure}[t]
\epsfxsize=190pt
\centerline{\epsffile{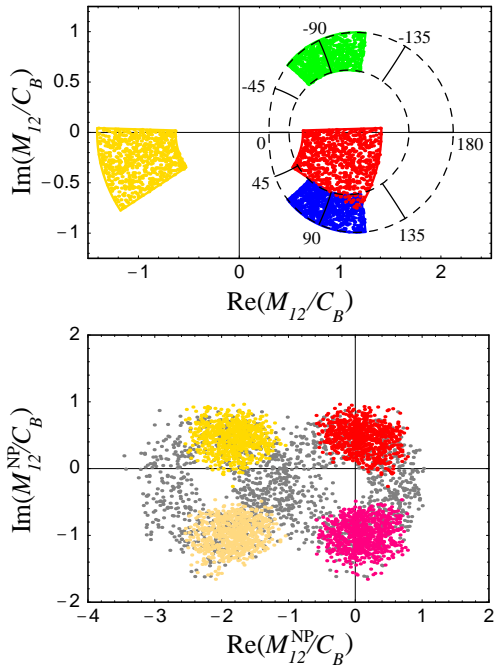}}
\caption{Determination of the $B_d$--$\bar B_d$ mixing amplitude 
$M_{12}$ (in units of $C_B$) in the complex plane, assuming 
present day uncertainties on the input parameters (see text). 
(a) Standard Model contribution $M_{12}^{\rm SM}$ 
(region bounded by the dashed circles) 
with marks indicating fixed values of $\gamma$. The filled regions 
between the circles correspond to $|\gamma|=(90\pm 25)^\circ$. The 
experimentally determined regions for $M_{12}$ are shown
for $\sin2\phi_d=0.26\pm 0.29$, where $2\phi_d\approx 15^\circ$ 
(middle right) and $2\phi_d\approx 165^\circ$ (left). 
(b) New Physics contribution $M_{12}^{\rm NP}$ corresponding to the
different regions: 
$(\gamma,2\phi_d)\approx(90^\circ,15^\circ)$ (upper right), 
$(\gamma,2\phi_d)\approx(-90^\circ,15^\circ)$ (lower right), 
$(\gamma,2\phi_d)\approx(90^\circ,165^\circ)$ (upper left), 
$(\gamma,2\phi_d)\approx(-90^\circ,165^\circ)$ (lower right). The 
rings of scatter points correspond to arbitrary $\gamma$.} 
\label{fig:NPnow}
\vspace*{-0.5cm}
\end{figure}

In Figure~\ref{fig:NPnow} we illustrate this analysis using 
present-day values of the input parameters from Ref.~\cite{CPRS00}
($|V_{ub}/V_{cb}|=0.085\pm 0.018$, 
$\sqrt{B_B} f_B=(0.21\pm 0.04)\,\mbox{GeV}$) and the average of the 
new BaBar 
and Belle results, $\sin2\phi_d=0.26\pm 0.29$. The upper plot 
shows the allowed regions for the Standard Model 
contribution to the mixing amplitude as well as for its experimental
value, taking into account the two-fold ambiguity in the mixing angle 
$2\phi_d$. The difference between any point in the Standard Model
regions with any point in the data regions defines an allowed vector 
in the complex $M_{12}^{\rm NP}$ plane. In the lower plot 
we show the resulting allowed regions 
for $M_{12}^{\rm NP}$. The origin in this plot 
corresponds to the Standard Model. We also show 
the results in the absence of any information on $\gamma$. 
An important message from this plot is that a potentially large
New Physics contribution (of order the Standard Model contribution) 
to the mixing amplitude is allowed by the 
data in large portions of parameter space. In order to find
out whether or not there is indeed such a large contribution it will
be necessary to determine $\gamma$ and resolve the discrete ambiguity
in $2\phi_d$.

\begin{figure}[t]
\epsfxsize=190pt
\centerline{\epsffile{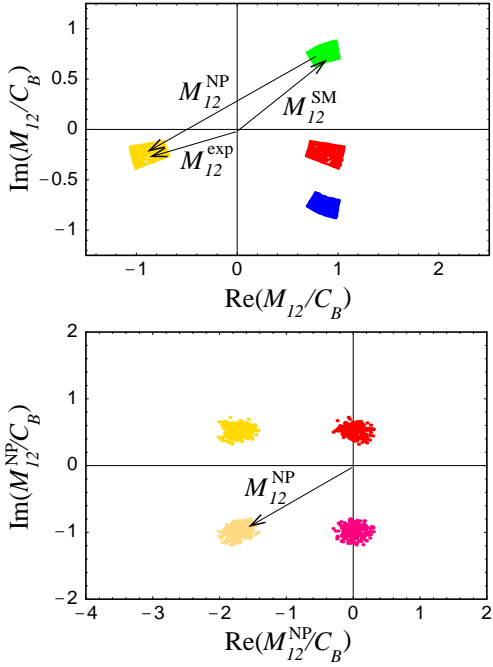}}
\caption{Same as Figure~\protect\ref{fig:NPnow}, but with smaller
uncertainties on input parameters. The arrows indicate the
construction of the New Physics contribution as explained in the text.}
\label{fig:NPfuture}
\vspace*{-0.5cm}
\end{figure}

Figure~\ref{fig:NPfuture} illustrates what the situation may look like
several years from now. By then the uncertainties in the input 
parameters will most likely have been reduced significantly, and the 
mixing angle will have been measured with good precision. For the
purpose of illustration we take $|V_{ub}/V_{cb}|=0.085\pm 0.009$, 
$\sqrt{B_B} f_B=(0.21\pm 0.02)\,\mbox{GeV}$, and 
$\sin2\phi_d=0.26\pm 0.10$. Also, the phase $\gamma$ will have been 
measured more accurately, perhaps with $\Delta\gamma=10^\circ$. This 
would lead to the picture shown in Figure~\ref{fig:NPfuture}(a) and 
to the allowed regions for New Physics shown in (b). Most importantly, 
as we have explained, in the long term the discrete ambiguities in 
both $\gamma$ and $2\phi_d$ will be removed, so that it would be 
possible to identify {\em which one of the four regions\/} in (b) 
is realized in nature. At this point, we will have achieved a
precise determination of the New Physics contribution to 
$B_d$--$\bar B_d$ mixing. 

\subsection{$B_s$--$\bar B_s$ mixing}

In the presence of New Physics in $B_d$--$\bar B_d$ mixing, it is 
very likely that also the $B_s$--$\bar B_s$ mixing amplitude is 
different from its value in the Standard Model. Therefore, 
measurements sensitive to this mixing amplitude should not be 
combined with measurements in the $B_d$ system. Rather, one should
probe for New Physics in the $B_s$ system in an independent way.
In the Standard Model the assumption of unitarity of the CKM matrix 
alone fixes the
magnitude of $|V_{tb}^* V_{ts}|$ (and hence the Standard Model 
contribution to $\Delta m_s$), and in addition implies that the 
$B_s$ mixing phase is very small, $\phi_s^{\rm SM}=O(\lambda^2)$
(with $\lambda\simeq 0.22$ the Wolfenstein parameter). Even at the
approved hadron collider experiments BTeV and LHCb it will not be
possible to measure this small Standard Model phase.
It follows that the complex amplitude 
$M_{12}^{\rm SM}(B_s)$ is determined by unitarity and is very nearly 
real. (In that sense the ``$B_s$ reference triangle'' is 
almost degenerate to a line.) 
Measuring the true values of the mass difference 
$\Delta m_s$ and of the mixing phase $2\phi_s$ (e.g., from 
the time-dependent CP asymmetry in $B_s\to J/\psi\,\phi$ decays), one 
can then construct the mixing amplitude from
$M_{12}(B_s)=(\Delta m_s/2)\,e^{-2i\phi_s}$.
The difference $M_{12}^{\rm NP}(B_s)=M_{12}(B_s)-M_{12}^{\rm SM}(B_s)$ 
determines directly the New Physics contribution to $B_s$--$\bar B_s$ 
mixing.

\subsection{$K$--$\bar K$ mixing}

The mass difference $\Delta m_K$ between the neutral 
kaon mass eigenstates is dominated by long-distance physics and does 
not admit a clean theoretical interpretation. Therefore, constraints 
on the CKM matrix from $K$--$\bar K$ mixing are derived only from the 
CP-violating quantity $|\epsilon_K|$, which (to a very good 
approximation) is given by 
$|\epsilon_K|\simeq|\mbox{Im}[M_{12}(K)]|/(\sqrt2 \Delta m_K)$. 
Consequently, one can only derive information on the New Physics 
contribution to the imaginary part of the mixing amplitude in 
the kaon system. 

The Standard Model contribution to $|\epsilon_K|$ is the product of a 
function of the Wolfenstein parameters 
$\bar\rho$ and $\bar\eta$ with a hadronic
matrix element parameterized by the quantity $B_K$ \cite{BBL96}. 
Once we have determined $\bar\rho$ and $\bar\eta$ (as a function of
$\gamma$ and $R_b$) from the reference
triangle, we can compute the Standard Model contribution. Subtracting
it from the measured value of $|\epsilon_K|$ gives the New Physics 
contribution, $|\epsilon_K^{\rm NP}|=|\mbox{Im}[M_{12}^{\rm NP}(K)]|/
(\sqrt2\Delta m_K)$.

\begin{figure}[t]
\epsfxsize=190pt
\centerline{\epsffile{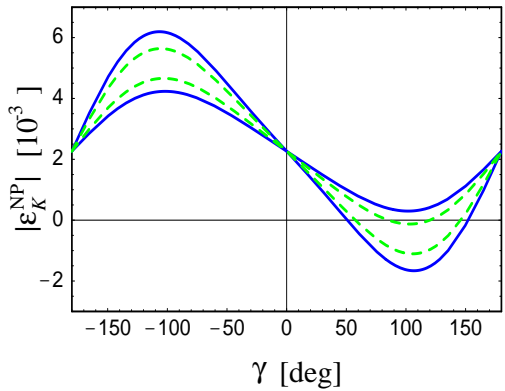}}
\caption{New Physics contribution to $|\epsilon_K|$ in units of 
$10^{-3}$, assuming present day uncertainties (region bounded by 
dark solid curves, corresponding to $B_K=0.86\pm 0.10$, 
$|V_{ub}/V_{cb}|=0.085\pm 0.018$) and future smaller errors 
(region bounded by light dashed curves, corresponding to 
$B_K=0.86\pm 0.05$, $|V_{ub}/V_{cb}|=0.085\pm 0.009$).}
\label{fig:NPkaon}
\vspace*{-0.5cm}
\end{figure}

In Figure~\ref{fig:NPkaon} we show the New Physics contribution to
$|\epsilon_K|$ as a function of $\gamma$, for both present-day and 
more long-term uncertainties on $B_K$ and $|V_{ub}/V_{cb}|$.
It is evident that a measurement
of $\gamma$ is the key ingredient needed to answer the question of
whether or not there is New Physics in $K$--$\bar K$ mixing. Once
$\gamma$ is known, $|\mbox{Im}[M_{12}^{\rm NP}(K)]|$ can be extracted
with good precision.

\section{New Physics in decays}
\label{sec:boxes}
\vspace{-0.1cm}

In order for a reference triangle construction to give the true value 
of $\gamma$, and thus be useful for extracting potential New Physics 
contributions to the mixing amplitudes, it must not be 
contaminated by additional New Physics effects in the associated 
decay amplitudes. Up to now we have assumed consistency between the 
different constructions, which would imply that to a good approximation 
New Physics enters only in the mixing amplitudes. This is indeed the 
case in many extensions of the Standard Model, particularly if the 
scale of new flavor interactions is at a TeV or beyond. Of 
course, it would be extremely exciting if the different reference 
triangle constructions were {\em not\/} consistent with one another, 
implying that there is New Physics in $B$ decay amplitudes, or that the 
3-generation CKM matrix is not unitary.

In Section~\ref{sec:RT} we have discussed various reference triangle 
constructions in roughly the chronological order in which it will be 
possible to carry them out. Interestingly, this order also corresponds 
to a progression from reliance on decays which in the Standard Model 
are penguin-dominated (and therefore more susceptible to New 
Physics) to those which are tree-dominated (and therefore less 
susceptible to New Physics), and finally to decays that are only based
on tree topologies. 

We briefly discuss tests for New Physics contributions to the 
penguin-dominated decays based on $b\to s\bar q q$ transitions in the
Standard Model. These tests can be carried out at various stages of 
data collection. The measurement of several CP-averaged
$B\to\pi K$ and $B\to\pi\pi$ decay rates itself provides for a series
of internal consistency checks. For instance, there are upper and 
lower bounds on certain rate ratios, which are based on flavor 
symmetries and rely on minimal theoretical input. Violation of these 
bounds would be a signal for new isospin-violating New Physics 
contributions \cite{GKN99}. In the longer term, as measurements of 
CP asymmetries for rare decays become available, additional tests 
will become possible. For example, one can then check whether the 
time-dependent CP asymmetries 
in $B\to J/\Psi\,K_S$ and $B\to\phi K_S$ decays are in agreement. A 
discrepancy would imply new CP-violating contributions 
to the $B\to\phi K_S$ decay
amplitude, which to a good approximation is a pure $b\to s\bar s s$ 
penguin amplitude in the Standard Model \cite{GW97}. 
There are also upper bounds on the direct CP asymmetries for 
$B^\pm\to\pi^\pm K^0$ and $B^\pm\to\phi K^\pm$ in the Standard Model, 
which could turn out to be violated \cite{BFM98,FKNP98}. 
Finally, a large direct CP asymmetry in $B^\pm\to X_s\gamma$ decays 
would imply significant New Physics contributions to $b\to s$ penguin
transitions \cite{KN98}.

If all penguin-dominated 
determinations of $\gamma$ are consistent, then 
it is unlikely that there is New Physics in decays, and the program 
we have outlined above for extracting New Physics in mixing can 
already be carried out with confidence. If these determinations are
not all consistent, however, the tree-dominated or pure tree
reference triangle constructions will be required in order to reliably
extract New Physics contributions to mixing. 
(At the same time, a clean determination of $\gamma$ would be
be required in order to obtain a detailed picture of the 
New Physics in the penguin-dominated decays \cite{GKN99}.)
The determination of $\gamma$
from the mixing-induced CP asymmetry in $B\to\pi^+\pi^-$ is less 
susceptible to New Physics effects, since these would have to compete 
with a Cabibbo-enhanced tree-level amplitude. Finally, the extractions 
of $\gamma$ from the pure tree processes $B\to D^{(*)}\pi$
and $B\to D K$ can only be affected by New Physics in rather exotic 
scenarios.  
Thus, checking for consistency between these two measurements
and the $B\to\pi^+\pi^-$ measurements essentially provides 
a test for New Physics effects in $B\to\pi\pi$.

If dedicated experiments to study the very rare $K\to\pi\nu\bar\nu$
decay modes will be performed, they will provide direct measurements
of the magnitude and phase of $V_{td}$ independent of mixing. 
The comparison of the kaon reference
triangle with the triangles obtained from $B$ decays primarily allows 
us to probe for New Physics in these rare kaon decays, which in the
Standard Model are mediated by box and electroweak penguin diagrams.
In addition, if the kaon and $B$-mesons triangles were to 
agree with one
another, this would be a direct test of the assumption of 3-generation 
CKM unitarity, as it would check the relation 
$V_{ub}^* V_{ud}+V_{cb}^* V_{cd}+V_{tb}^* V_{td}=0$ independent of
any mixing measurements. 
Another test of CKM unitarity would be the direct measurement of
the element $|V_{tb}|$ (and perhaps $|V_{ts}|$) in top decay.

\vspace{-0.1cm}
\section{Conclusions}
\vspace{-0.1cm}

Motivated by today's 
announcement of the first $\sin2\beta_{\psi K}$ 
measurements from the dedicated $B$-factory experiments BaBar and 
Belle, we have reconsidered strategies for exploring New Physics in
$B$--$\bar B$ and $K$--$\bar K$ mixing in a model-independent way.
The low central values found by these experiments raise the 
possibility that there is New Physics in $B_d$--$\bar B_d$ mixing. 
Therefore it becomes 
crucial to base studies of flavor physics on comparisons with a 
reference unitarity triangle whose construction is 
independent of mixing measurements. Ultimately, such a strategy is 
preferable whether or not the measured value of 
$\sin2\beta_{\psi K}$ agrees with the prediction 
from the global analysis of the unitarity triangle. 

We have described in detail a program that in a few years could 
cleanly determine the New Physics contributions (in magnitude and 
phase) to the $B_d$--$\bar B_d$ and $K$--$\bar K$ mixing amplitudes. 
(A similar, more straightforward analysis for $B_s$--$\bar B_s$ 
mixing can be performed at the BTeV and LHCb experiments.) Similar 
strategies have been proposed previously by several authors. Here we 
have stressed the relevance of the progress recently made in devising 
strategies for near-term measurements of the weak phase $\gamma$ 
based on charmless hadronic $B$ decays. Knowing $\gamma$ with 
reasonable accuracy, and without discrete ambiguities, is {\em the\/} 
key element that makes the program outlined in this Letter feasible 
and very powerful. We have also pointed out that the comparison of 
different constructions of the reference triangle provides several 
opportunities for probing New Physics in decay amplitudes, not related 
to mixing. Nothing would be more exciting than to follow the 
unfolding of New Physics at the $B$ factories in the next few years
ahead of us.

{\it Acknowledgements:\/} 
We wish to thank the SLAC Theory Group for its hospitality while this
work was carried out.
A.K.\ is supported by the Department of Energy under Grant 
No.~DE-FG02-84ER40153. M.N.\ is supported in part by the National 
Science Foundation.

\end{document}